\documentclass[11pt,a4paper]{article}
\usepackage{jcappub}
\usepackage{amssymb}
\usepackage{graphicx}
\usepackage{amsmath}
\usepackage{amsfonts}
\usepackage{hyperref}

\title{Maximum mass of stable magnetized highly super-Chandrasekhar white dwarfs: stable solutions with
varying magnetic fields}

\author[a]{Upasana Das}
\author[a]{and Banibrata Mukhopadhyay \footnote{Corresponding Author.}}

\affiliation[a]{Department of Physics, Indian Institute of Science, 
Bangalore 560012, India}

\emailAdd{upasana@physics.iisc.ernet.in}
\emailAdd{bm@physics.iisc.ernet.in}

\abstract{We address the issue of stability of recently proposed significantly super-Chandrasekhar white dwarfs.
We present stable solutions of magnetostatic equilibrium models for super-Chandrasekhar white dwarfs 
pertaining to various magnetic field profiles. This has been obtained 
by self-consistently including the effects of the magnetic pressure gradient and total magnetic density
in a general relativistic framework. We estimate that the maximum stable mass of magnetized white dwarfs
could be more than $3$ solar mass. This is very useful to explain peculiar, overluminous type~Ia
supernovae which do not conform to the traditional Chandrasekhar mass-limit.}


\arxivnumber{1404.7627}

\begin{document}
\maketitle


\section{Introduction}


Type~Ia supernovae are one of the most widely studied astronomical events and rightfully so because of 
their enormous importance in measuring cosmic distances. Although not everything is understood about 
these events, the general consensus is that they are produced due to the thermonuclear explosion of a 
white dwarf having mass very close to the Chandrasekhar limit of $1.44M_\odot$, when $M_\odot$ 
being the mass of Sun \cite{chandra35}. 
However, recent observations of a fast-growing number of several peculiar, 
overluminous type~Ia supernovae, e.g. SN~2006gz,
SN~2007if, SN~2009dc, SN~2003fg, bring even this basic idea into serious question, as they are best 
explained by invoking progenitor white dwarfs with super-Chandrasekhar mass in the range 
$2.1-2.8M_\odot$ \cite{nature,scalzo,hicken,yam,silverman,
taub}. 

In our previous works \cite{kundu,prd12,ijmpd12,prl13,apjl13,grf13,mpla14}, we addressed this issue 
and showed that strongly magnetized white dwarfs can violate the sacrosanct Chandrasekhar limit significantly, 
thus solving the puzzle posed by these peculiar supernovae. We essentially modeled the central region of 
the white dwarf having a strong constant magnetic field $B$, which would fall off near the surface. 
However, a few authors \cite{dong,chamel} have questioned the stability 
of our super-Chandrasekhar white dwarfs, to which we have partially responded in our latest work \cite{mpla14} and 
which we address more rigorously and quantitatively in the current work. 
On the other hand, Federbush et al. \cite{smoller} have carried out an extensive mathematical analysis 
of stable magnetic star solutions including the polytrope describing our super-Chandrasekhar white dwarfs
\cite{prl13}. They have proved that such a polytrope will remain stable, provided the underlying 
magnetic field profile is constrained appropriately.

We take this opportunity to point out that 
one of the authors, Dong et al. \cite{dong}, have argued completely erroneously that our model of 
super-Chandrasekhar white dwarfs would lead to an unphysical large mass, if the contribution of 
magnetic density, $\rho_B=B^2/(8\pi c^2)$, is included.
They have incorrectly calculated a value of $24 M_\odot$ as the contribution of 
$\rho_B$ to the total mass of the particular white dwarf having {\it central} 
magnetic field $B_{\rm cent}=8.8\times 10^{17}$ G, 
radius $R=70$ km and mass $M=2.58 M_\odot$ (computed without $\rho_B$) \cite{prl13}. 
If, following Dong et al., we indeed consider a constant 
$B$ throughout the above white dwarf, then to arrive at the $M$ and $R$ correctly in the Newtonian framework, one must solve the 
following set of equations simultaneously (when magnetic pressure gradient $dP_B/dr=0$
and magnetic tension
$\vec{B}.\nabla\vec{B}|_r=0$, as chosen by Dong et al. and us previously):
\begin{equation}
\frac{dP(r)}{dr} + \frac{dP_B}{dr} = -\frac{G M(r)}{r^2} (\rho(r) + \rho_B)+\frac{\vec{B}\cdot\nabla\vec{B}|_r}
{4\pi},\,\,\,\,\frac{dM(r)}{dr}=4\pi r^2(\rho(r)+\rho_B),
\end{equation}
where $r$ is the radial distance from the center of the white dwarf, 
$P$ the matter pressure, $P_B$ the magnetic pressure, $\rho$ the matter density and $G$ Newton's gravitation constant.
Mistake of Dong et al was {\it not} considering $\rho_B$ in the first equation but considering {\it only} in the second equation 
above. A self-consistent inclusion of $\rho_B$, however, shows that both $M$ and $R$ of the 
white dwarf decrease drastically, due to the increased gravitational force because of the increased total density
(see Figure 1 and \S5 of \cite{mpla14}). 
When $B$ decreases away from the center, 
one must additionally account for the effects of $dP_B/dr$ (and $\vec{B}.\nabla\vec{B}|_r$), which will lead to 
an increase of $M$ and $R$, as done in the current work. 
The point to be emphasized is that if one wants to include the effects of $P_B$ and $\rho_B$ 
self-consistently, then they have to be included in both the equations above, which unfortunately 
Dong et al. did not do. To avoid complicacy, we, earlier, did not include these 
effects in either of the equations. We shall show here that our earlier solutions \cite{prl13} are 
much closer to the exact results compared to that obtained by Dong et al \cite{dong}.

In this work, 
we present stable, magnetostatic equilibrium solutions of significantly super-Chandrasekhar white dwarfs in the framework of general 
relativity, obtained
by self-consistently including the effects of the pressure gradient due to a varying $B$ 
and also $\rho_B$. 
Note that a general relativistic approach is important in order to accurately understand 
the effects of $\rho_B$ and $P_B$ for the high density and high $B$ white dwarfs.
The current work firmly 
corroborates the existence of our super-Chandrasekhar white dwarfs.

In the next section, we discuss the constraints imposed on the magnetic field profiles along with the 
equations to be solved. Subsequently, we describe various solutions of super-Chandrasekhar 
white dwarfs in section 3. Finally, we end with conclusions in section 4.


\section{Magnetic field profiles and basic equations}


Since magnetized white dwarfs form the basis of our work, it is important to note that 
a large number of them have been discovered by the Sloan Digital Sky Survey (SDSS), 
having high surface fields $10^5-10^9$G \cite{schmidt03,vanlandingham05}. 
It is likely that the observed surface field is several orders of magnitude smaller 
than the central field. Thus, it is important to perform numerical calculations, in the presence 
of a varying $B$, to self-consistently account for $dP_B/dr$ and $\rho_B$, 
in addition to $dP/dr$. 
Keeping this in mind, we model the variation of $B$ as a function of $\rho$ 
inside the white dwarf by adopting the profile proposed by Bandyopadhyay et al. \cite{bando}. This 
is very commonly applied to magnetized neutron stars (see, e.g., \cite{sinha}) given by

\begin{equation}
B \left(\frac{\rho}{\rho_0}\right) = B_s + B_0\left[1-\exp \left(-\eta \left(\frac{\rho}{\rho_0}\right)^\gamma \right)\right],
\label{bprofile}
\end{equation}
where $\rho_0$, for the present purpose, is chosen to be one-tenth of the {\it central} matter density ($\rho_c$) 
of the corresponding white dwarf, 
$B_s$ is the surface magnetic field, $B_0$ is a parameter having 
dimension of $B$, $\eta$ and $\gamma$ are dimensionless parameters which determine how exactly the 
field decays from the center to the surface. Note that as $\rho \rightarrow 0$ close to the surface of the white dwarf, $B \rightarrow B_s$. 

In this work, we present the cases with $10^9$ G $\leq B_s \leq 10^{12}$ G. Note that the lower limit 
for $B_s$ is guided by the aforementioned SDSS observations. 
However, for $B_{\rm cent}\geq 10^{14}$ G, what we will consider in this work, 
the result is independent of $B_s$ chosen above.
Now, it has already been 
shown in our previous works \cite{prd12,prl13} that for $B>B_{cr}=4.414\times 10^{13}$ G, the effect of Landau quantization 
becomes important which modifies the equation of state (EoS) of the electron degenerate matter constituting 
the white dwarf and gives rise to super-Chandrasekhar white dwarfs. Hence, in this work we consider 
different values of $B_0$ such that the resulting 
$B_{\rm cent}$ of the white dwarf is super-critical, i.e. $B_{\rm cent}> B_{cr}$. 
We recall that the maximum number of Landau levels ($\nu_m$) occupied by the electrons is given by \cite{prd12}
\begin{equation}
\nu_m= \frac{\left(\frac{E_{Fmax}}{m_ec^2}\right)^{2} - 1}{2B/B_{cr}} ,
\label{numax}
\end{equation}
where $E_{Fmax}$ is the maximum Fermi energy of the system, $m_e$ the mass of the electron 
and $c$ the speed of light.
Thus, for a fixed $E_{Fmax}$, as $B$ 
decreases from the center to the surface of the white dwarf, 
$\nu_m$ increases. Consequently, the underlying EoS is constructed taking this variation 
of $B$ into account, such that near the center of the white dwarf, it is (strongly) Landau 
quantized (see, e.g., \cite{prd12}), while away from the center it approaches Chandrasekhar's EoS \cite{chandra35}.

Now, it is known that in the presence of a strong $B$, the total pressure 
of the system may 
become anisotropic \cite{sinha}, such that it is in the direction perpendicular to $B$ 
given by $P_{\perp}=P+B^2/(8\pi)$ (neglecting magnetization, which is much smaller compared 
to $B^2/(8\pi)$ for the $B$ of present interest), while that in the direction parallel to 
$B$ is given by $P_{||}=P-B^2/(8\pi)$. 
If $B$ becomes too large, 
then one notices that 
$P_{||}$ may vanish (and maybe even negative) and this might lead to instabilities in the system. 
For the present purpose, however, we model our white dwarfs with varying $B$, by adopting the same magnetostatic 
balance equation as in general relativity, 
well known as the (magnetized) Tolman-Oppenheimer-Volkoff (TOV) equation \cite{wald} (which is applicable 
for compact objects having an overall isotropic pressure), using isotropic magnetic
pressure to be $P_B=B^2/(24\pi)$. 
Validity of such a consideration has been already justified earlier
\cite{adam,mpla14,cheoun,herrera,herrera2}. In particular, Cheoun et al. \cite{cheoun}
invoked randomly oriented magnetic domains, where the random
currents cancel each other, to justify the spherically
symmetric magnetic field configuration of the strongly magnetized neutron stars.
This corresponds to force free fields in the domain forming region.
On the other hand, Herrera \& Barreto \cite{herrera2} 
invoked a general formalism to describe general relativistic polytropes for anisotropic 
matter, showing the possibility of two different polytropic EoSs.  
The present work can be treated as an application of this formalism, 
when the density of the polytropic EoS
consists only of the matter density (their Case I) such that $P=K\rho^{1+1/n}$, 
corresponding to a particular case of the
polytropes, where $n$ is the polytropic index and $K$ the polytropic constant. 
Furthermore, the white dwarfs in the current work are the polytropes for anisotropic 
fluids described by Herrera \& Barreto \cite{herrera2}, with appropriate modification 
of the anisotropic contribution appearing in their Eq. (53) by the magnetic pressure and density.
Although they noted that spherical symmetry can be broken in the presence of
a very strong magnetic field, when magnetized Fermi fluids may be regarded as anisotropic
systems, the choice of spherical white dwarfs in the present work, however, is not expected 
to change the main conclusions. Hence, the set of equations we solve are 
\begin{equation}
\frac{dM(r)}{dr} = 4\pi r^2 (\rho(r)+\rho_B),
\label{tov_mass}
\end{equation}
\begin{equation}
\frac{d\rho(r)}{dr} = -\frac{G\left(\rho(r)+\rho_B + \frac{(P(r)+P_B)}{c^2}\right)\left(M(r)+ \frac{4\pi r^3 (P(r)+P_B)}{c^2}\right)}{r^2\left(1-\frac{2GM(r)}{c^2r}\right)\left(\frac{dP(r)}{d\rho} +\frac{dP_B}{d\rho}\right)} .
\label{tov_dens}
\end{equation} 

\begin{figure*}
\begin{center}
\includegraphics[angle=0,width=18cm]{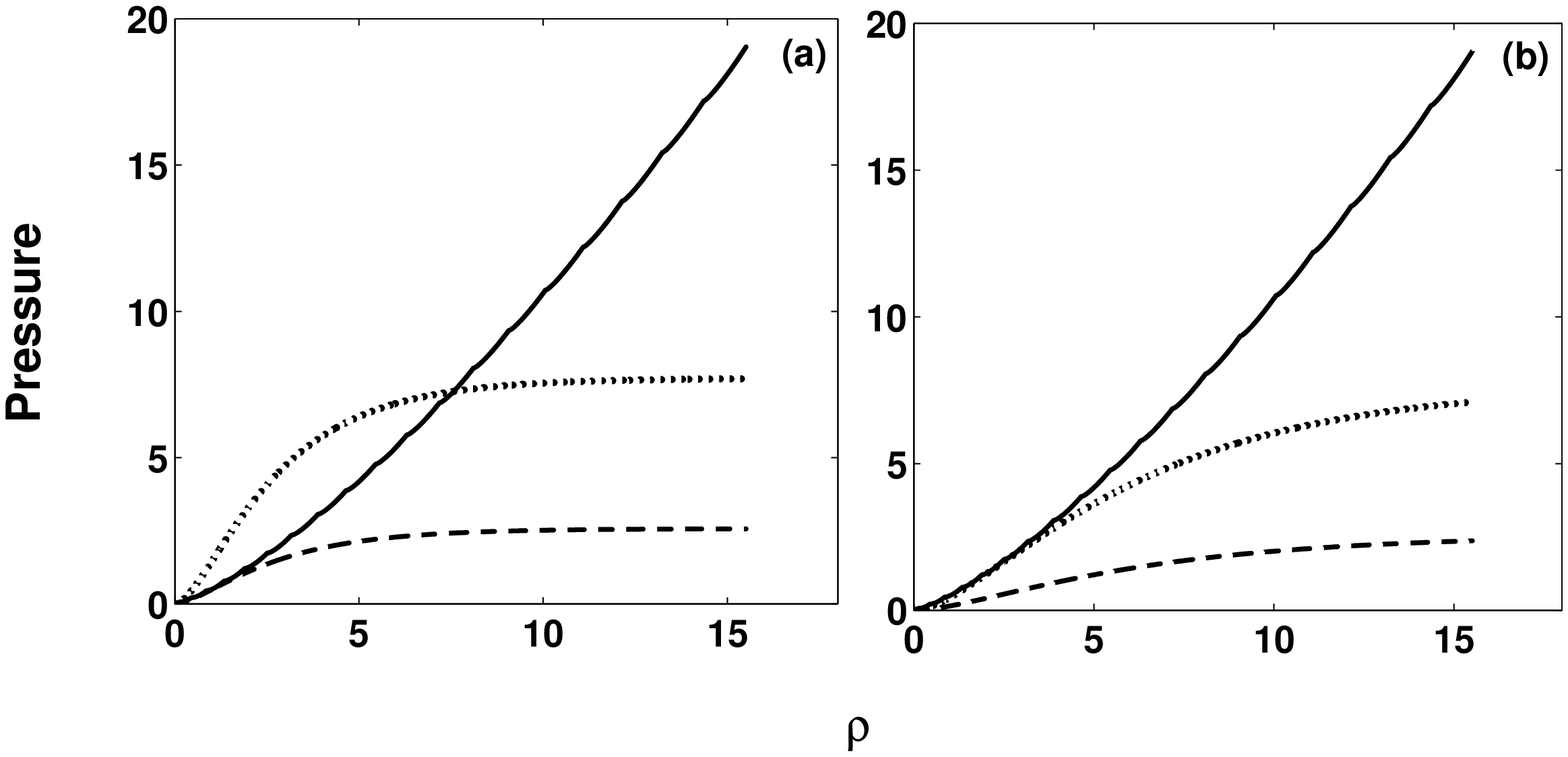}
\caption{ Various pressures as functions of $\rho$ --- (a) $P\geq B^2/(24\pi)$ throughout (constraint (i)).  
(b) $P\geq B^2/(8\pi)$ throughout (constraint (ii)). The 
solid, dotted and dashed lines represent $P$, 
$B^2/(8\pi)$ and $B^2/(24\pi)$ respectively. Pressure is in units of $10^{27}$ ergs/cc and $\rho$ in 
units of $10^9$ gm/cc. 
  }
\label{pressure}
\end{center}
\end{figure*}

Keeping the above consideration of a possible instability due to anisotropic pressure in mind, we 
propose two plausible constraints on 
the magnetic field profile in this work, such that: (i) the average parallel 
pressure, given by $P-P_B$, should remain positive throughout the white dwarf, (ii) the parallel 
pressure given by $P_{||}$ should remain positive throughout the white dwarf (note that (ii) 
automatically guarantees (i) as well). The two constraints have been illustrated in 
Figure \ref{pressure} with an example. Note that an attempt to model magnetized,
significantly super-Chandrasekhar white dwarfs was made earlier in the Newtonian framework,
however, without imposing realistic constraints on the field \cite{adam}.

Figure \ref{pressure}(a) represents constraint (i) for a given EoS.
Here, $B_{\rm cent}=4.4\times 10^{14}$ G, $E_{Fmax}=20m_e c^2$, the number of occupied Landau levels 
at the {\it center} $\nu_{mc}=20$ (as calculated from Eq. (\ref{numax})), $\gamma=0.9$ and $\eta=0.843$. 
Note that Figure \ref{pressure}(a) represents a limiting magnetic profile, such that $P_B$ is just equal to $P$ at some 
point inside the white dwarf. For a fixed $\gamma$, if one varies $\eta$, then one obtains such a limiting 
profile for a maximum value of $\eta=\eta_{max}$. Only for $\eta<\eta_{max}$, $P_B<P$ throughout.
Similarly, Figure \ref{pressure}(b) represents constraint (ii) for $B_{\rm cent}=4.22\times10^{14}$ G, $E_{Fmax}=20m_ec^2$, 
$\nu_{mc}=21$, $\gamma=0.9$ and $\eta_{max}=0.403$. 
Note that the significance of the limiting cases shown in 
Figures \ref{pressure}(a) and (b) lie in the fact that they correspond to the maximum stable mass of the white dwarf in each case, 
in the presence of a varying $B$, as we shall see next. 

\begin{figure*}
\begin{center}
\includegraphics[angle=0,width=18cm]{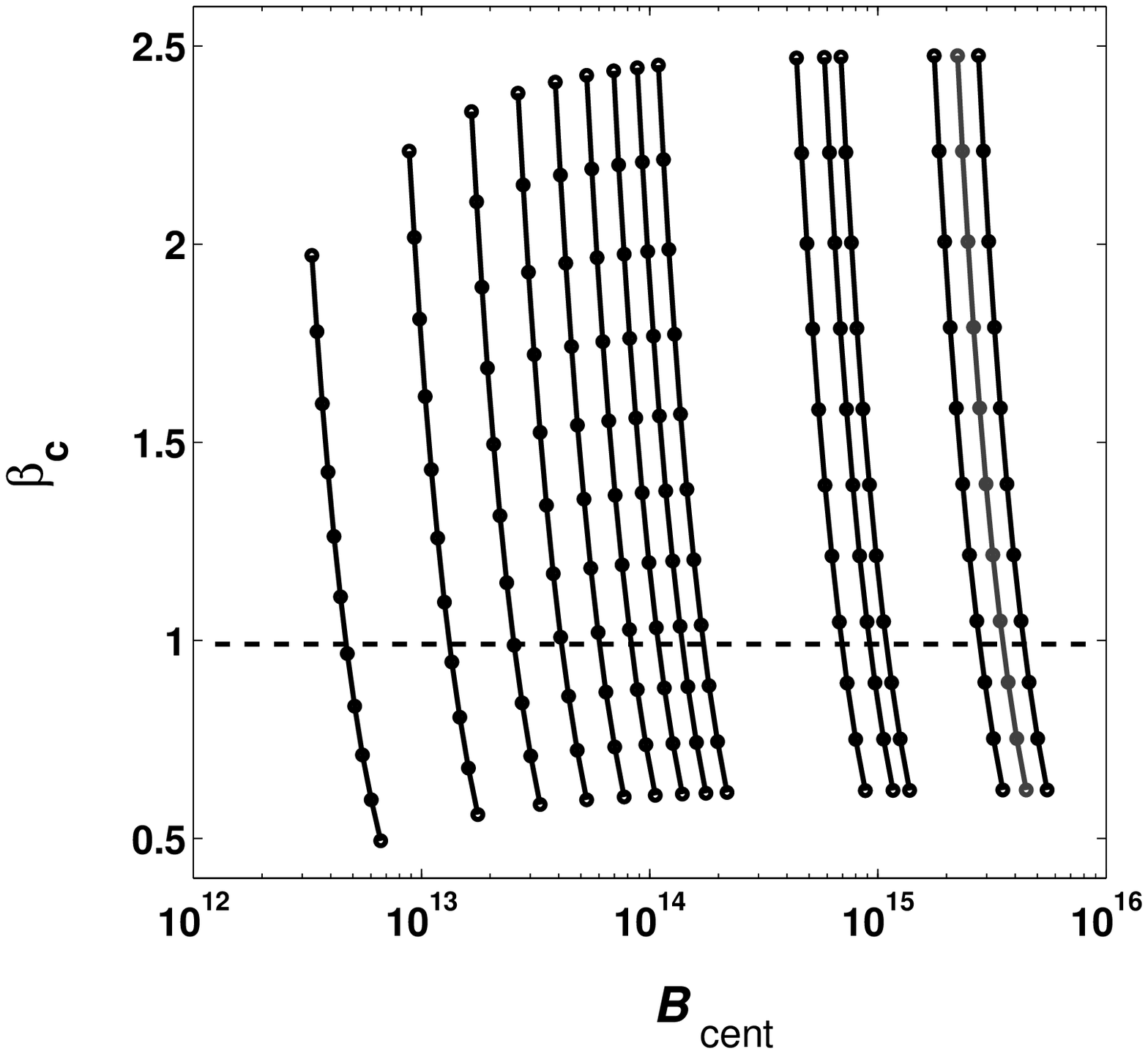}
\caption{ Plasma-$\beta_c$ as a function of $B_{\rm cent}$ for various EoSs --- from left to right each solid line corresponds to $E_{Fmax}=2,3,4,5,6,7,8,9,10,20,23,25,40,45$ 
and $50m_ec^2$ respectively. The filled circles on each solid line correspond to, 
from top to bottom, $\nu_{mc}=20,19,18,17,16,15,14,13,12,11$ and $10$ respectively. $B_{\rm cent}$ is in units of G.
  }
\label{betac}
\end{center}
\end{figure*}


\section{Solutions of super-Chandrasekhar white dwarfs}


\begin{figure*}
\begin{center}
\includegraphics[angle=0,width=18cm]{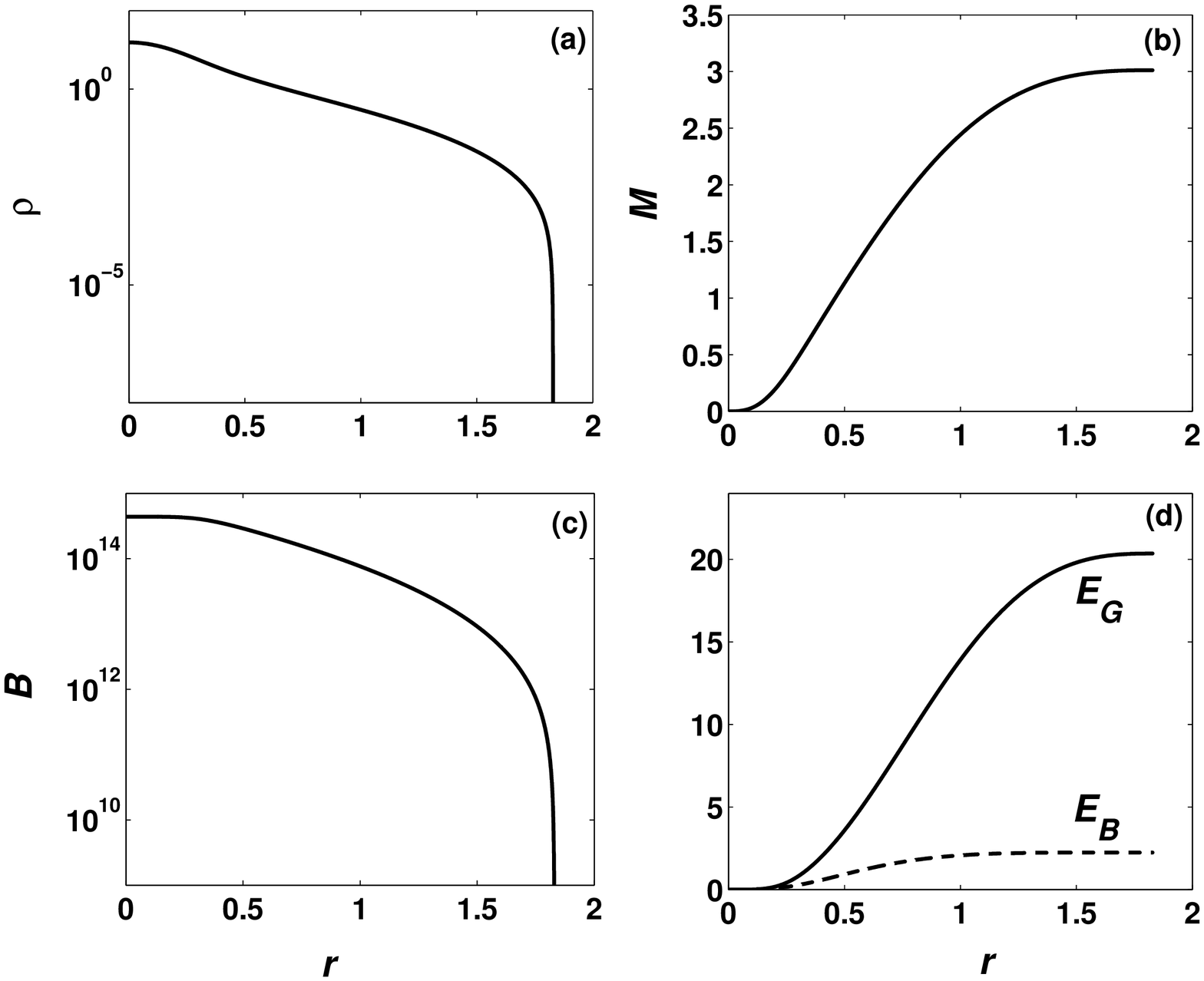}
\caption{ Solutions of a super-Chandrasekhar white dwarf having EoS corresponding to Figure \ref{pressure}(a). 
(a) $\rho$, 
(b) $M$, (c) $B$,  as functions of $r$ within the white dwarf. (d)  
$E_G$ and $E_B$ as functions of $r$. $M$, $\rho$, $r$ and $B$ are in 
units of $M_\odot$, $10^9$ gm/cc, 1000 km and G respectively, while $E_G$ and $E_B$ 
are in units of $10^{51}$ ergs.
  }
\label{WDin}
\end{center}
\end{figure*}

With the magnetic profiles set according to the above constraints, 
we now solve Eqs. (\ref{tov_mass}) and (\ref{tov_dens}) subjected to the boundary conditions 
$\rho(0)=\rho_c$ and $M(0)=0$.
As an additional criterion for stability, independent of 
constraints (i) and (ii), $\rho_c$ is chosen such that the 
plasma-$\beta$ at the center, $\beta_c=P_c/(B_{\rm cent}^2/8\pi) > 1$ (i.e., {\it central} $P_{||}>0$) 
always, where $P_c$ is the {\it central} matter pressure.
If $P_c$ is the maximum possible pressure for a given EoS, which is determined by the choice of 
$E_{Fmax}$ and $\nu_{mc}$ (see Eq. (17) of \cite{prd12} for the expression of $P$ for a Landau quantized system), 
then we arrive at a very interesting result as illustrated in Figure \ref{betac}. 
In Figure \ref{betac}, we compute $\beta_c$ for different combinations of $E_{Fmax}$ and $\nu_{mc}$, such that 
only the region above the $\beta_c=1$ line is allowed as per our condition. One observes that 
as $E_{Fmax}$ increases from 2 to $6m_ec^2$, the minimum allowed value of $\nu_{mc}$, 
for which $\beta_c \geq 1$ {\it just}, decreases from 15 to 13. However, for $E_{Fmax}\gtrsim 6 m_e c^2$, the minimum 
allowed value of $\nu_{mc}$ saturates at $13$, i.e., $\beta_c>1$ is satisfied 
for all $B_{\rm cent}$ having $\nu_{mc}\geq13$. 
This additionally constrains the maximum possible $B_{\rm cent}$ for a given $E_{Fmax}$, 
which also corresponds to the maximum stable mass of the white dwarf. 

Figure \ref{WDin} depicts the solutions of Eqs. (\ref{tov_mass}) and (\ref{tov_dens}) for 
a single white dwarf, having $\rho_c=1.55\times 10^{10}$ gm/cc and $B_{\rm cent}=4.4\times 10^{14}$ G, 
whose EoS and necessary magnetic field constraints are portrayed in Figure \ref{pressure}(a). 
Figures \ref{WDin}(a), (b) and (c) show the variations of $\rho$, $M$ and $B$ respectively 
inside the white dwarf. The resulting white dwarf is clearly highly super-Chandrasekhar, having $M=3.01M_\odot$ 
and $R=1828$ km. Figure \ref{WDin}(d) shows the variations of the gravitational energy
$E_G$ and the magnetic energy $E_B$ inside the white dwarf. 
The total gravitational energy of the white dwarf is given by $E_{GT}=\int_0^R \phi~ dM$, 
where $\phi$ is the gravitational potential in general relativity including magnetic effects, while the 
total magnetic energy of the white dwarf is given by $E_{BT}=\int_0^R P_B~ d^3r$. Very importantly, we 
observe that for this white dwarf, $E_{GT}=2.035\times 10^{52}$ ergs, which is almost an order of 
magnitude larger than $E_{BT}= 2.246\times 10^{51}$ ergs, thus firmly establishing it as stable. 

\begin{figure*}
\begin{center}
\includegraphics[angle=0,width=18cm]{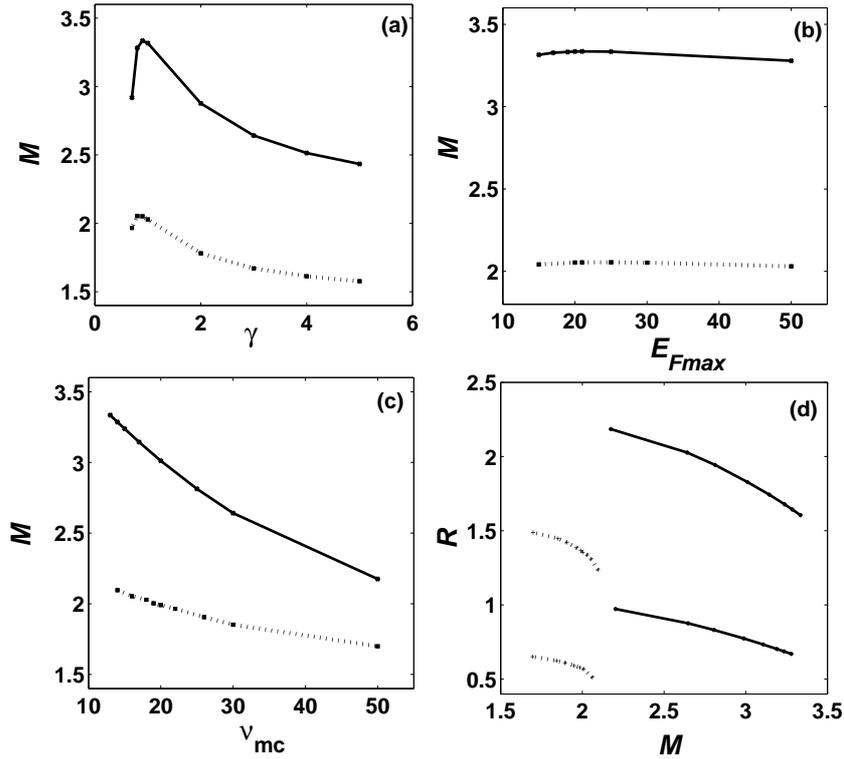}
\caption{ Super-Chandrasekhar white dwarfs with varying $B$ --- The solid and dotted lines 
represent the cases corresponding to constraints (i) and (ii) respectively (see text). 
(a) $M$ as a function of $\gamma$ for $E_{Fmax}=20m_ec^2$, $B_{\rm cent}=6.77\times 10^{14}$ G, 
$\eta=\eta_{max}$ for respective $\gamma$s. (b) $M$ as a function of $E_{Fmax}$ for $B_{\rm cent}=6.77\times 10^{14}$ G, 
$\gamma=0.9$ for solid line and $B_{\rm cent}=5.18\times 10^{14}$ G, $\gamma=0.8$ for dotted line. (c) $M$ as a function 
of $\nu_{mc}$ for $E_{Fmax}=20m_ec^2$, $\gamma=0.9$ for solid line and $\gamma=0.8$ for dotted line. 
(d) The topmost solid and dotted lines represent the $M$-$R$ relations corresponding to (c), while the solid and 
dotted lines at the bottom represent the $M$-$R$ relations corresponding to (c) but with $E_{Fmax}=50m_ec^2$. 
$M$, $R$ and $E_{Fmax}$ are in units of $M_\odot$, 1000 km and $m_ec^2$ respectively.
  }
\label{mass}
\end{center}
\end{figure*}

Various results obtained by considering different magnetic field profiles, $\nu_{mc}$ and $E_{Fmax}$,
are summarized in Figure \ref{mass}. We further restrict $E_{Fmax}$ to $50m_ec^2$, in order to avoid possible
neutronization of the matter.
Figure \ref{mass}(a) shows the variation of $M$ with $\gamma$ for a fixed
$E_{Fmax}$ and $B_{\rm cent}$. 
We note that as $\gamma$ increases, $M$ also increases, until $\gamma=0.9$ for constraint (i) 
and $\gamma=0.8$ for constraint (ii), after which $M$ starts 
decreasing again. The maximum mass ($M_{max}$) for constraint (i) is $3.33M_\odot$ while that for 
constraint (ii) is $2.05M_\odot$.

Figure \ref{mass}(b) shows the variation of $M$ with $E_{Fmax}$ for a fixed $B_{\rm cent}$ and $\gamma$. 
Since we are interested in the maximum possible mass of the white dwarf, we fix 
$\gamma$ accordingly. 
We note that as $E_{Fmax}$ increases, $M$ attains 
a peak and then decreases very slightly. For constraint (i) $M_{max}= 3.33M_\odot$ at $E_{Fmax}\sim 21 m_ec^2$, 
while for constraint (ii) $M_{max}= 2.05M_\odot$ at $E_{Fmax}\sim 25 m_e c^2$.

Figure \ref{mass}(c) shows the variation of $M$ with $\nu_{mc}$ for a fixed $E_{Fmax}$ and $\gamma$. We observe that 
with the decrease in $\nu_{mc}$ or a corresponding increase in 
$B_{\rm cent}$, $M$ increases, as was also shown in our 
previous works \cite{prd12,prl13}. For constraint (i) $M_{max}= 3.33M_\odot$, 
while for constraint (ii) $M_{max}=2.1M_\odot$, both at $B_{\rm cent}=6.77\times 10^{14}$ G (or $\nu_{mc}=13$).

Finally, Figure \ref{mass}(d) shows the mass-radius relations for the cases in 
Figure \ref{mass}(c) along with the corresponding cases with $E_{Fmax}=50m_ec^2$.
We observe that for 
both constraints (i) and (ii), $R$ decreases as $M$ increases, i.e. a stronger magnetic 
field makes the star more compact (as argued previously, see, e.g., \cite{prd12,prl13}). 
Additionally, we note that the radii of the 
white dwarfs having $E_{Fmax}=50m_ec^2$ are more than a factor of two smaller than those with $E_{Fmax}=20m_ec^2$ for 
roughly the same range of $M$. This is expected because a higher $E_{Fmax}$ implies a 
higher $\rho_c$ and hence more compact objects. For example, for the cases pertaining to constraint (i),
$M_{max}=3.33M_\odot$ and $R=1605$ km for $E_{Fmax}=20m_ec^2$ ($\rho_c=1.55\times 10^{10}$ gm/cc), 
while $M_{max}=3.28 M_\odot$ and $R=670$ km for $E_{Fmax}=50m_ec^2$ ($\rho_c=2.42\times 10^{11}$ gm/cc). 
Similarly, for the cases pertaining to constraint (ii), $M_{max}=2.1M_\odot$ and $R=1237$ km for $E_{Fmax}=20m_ec^2$,
while $M_{max}=2.06 M_\odot$ and $R=512$ km for $E_{Fmax}=50m_ec^2$.

In order to confirm the stability of these super-Chandrasekhar white dwarfs, we resort to 
Table \ref{table1}, which describes the properties of some of the super-Chandrasekhar white dwarfs having 
$E_{Fmax}=20m_ec^2$ shown in Figure \ref{mass}(d). The seven
columns represent respectively: the constraint conditions (i) or (ii) on the magnetic profile; 
$M$; $R$; $E_{GT}$; $E_{BT}$; 
the equipartition magnetic field $B_{\rm equi}$ as defined in \S2 of \cite{mpla14} (with 
polytropic index $n\approx 3$, as is true on average for the 
EoSs of the cases described in Table \ref{table1}); and the average magnetic field, $B_{\rm avg}=\frac{1}{R}\int_0^R B(\rho(r))~dr$, 
of the white dwarf. It is evident from Table \ref{table1} that for all the cases, 
$E_{GT}$ significantly dominates over $E_{BT}$, which assures the stability of any 
magnetized compact object. Although $B_{\rm equi}$ is only a rough estimate, we still 
note that $B_{\rm avg}$ in all the cases is substantially less than $B_{\rm equi}$, 
thus ensuring that the corresponding super-Chandrasekhar white dwarfs are highly stable.

\section{Conclusions}

\begin{table}[tbp]
\footnotesize
\centering
\renewcommand{\arraystretch}{1.4}
\begin{tabular}{|c|c|c|c|c|c|c|}
\hline
Constraint & $M$ ($M_\odot$) &  $R$ (1000 km) & $E_{GT}$ ($10^{51}$ ergs) & $E_{BT}$ ($10^{51}$ ergs) & $B_{\rm equi}$ ($10^{14}$ G) & $B_{\rm avg}$ ($10^{14}$ G) \\
\hline
& 2.17 & 2.18 & 12.75 & 0.77 & 1.22 & 0.72  \\
 (i) $P\geq B^2/(24\pi)$  & 2.64 & 2.03 &  17.01 & 1.46 & 1.72 & 1.13 \\
& 3.33 & 1.60 &  23.41 & 3.22 & 3.47 & 2.41 \\ \hline
& 1.70 & 1.49 & 7.99 & 0.30 & 2.06 & 0.76 \\
 (ii) $P\geq B^2/(8\pi)$ & 1.96 & 1.38 &  9.94 & 0.67 & 2.74 & 1.41 \\
& 2.10 & 1.24 & 10.88 & 0.93 & 3.67 & 2.04  \\ \hline
\end{tabular}
\caption{Properties of super-Chandrasekhar white dwarfs with $E_{Fmax}=20m_ec^2$ in Figure \ref{mass}(d).}
\label{table1}
\end{table}

We have reestablished the existence of stable, significantly super-Chandrasekhar white dwarfs 
by considering different, varying magnetic field profiles in a general relativistic framework, subjected to 
certain constraints to ensure stability. The maximum mass of the white dwarf can be as large as 
$3.33M_\odot$, depending on the underlying nature of the variation of $B$ with $\rho$. 
Note that a violation of the constraints on the magnetic field profile leads to unphysical 
white dwarfs. For example, for the case shown in Figure \ref{pressure}(a), if $\eta$ is chosen to be 3.5, 
which violates constraint (i), we obtain an unusually massive white dwarf having 
$M=28.4M_\odot$ and $R=6880$ km (note that even relatively smaller mass white dwarfs also 
could be unphysical if constraint (i) is violated).

For all the cases 
described in this work, $E_{GT} \gg E_{BT}$ and also $B_{\rm equi}>B_{\rm avg}$, which show that 
these magnetized super-Chandrasekhar white dwarfs are highly bound systems. We note that the 
same calculations, when carried out in a Newtonian framework, also yield stable white dwarfs, but with a 
slightly higher $M$ and $R$ for each of the cases discussed here. 

Note that in our previous works \cite{prd12,prl13}, we carried out spherically symmetric 
calculations in the Newtonian framework, 
by assuming that the field is constant in a large central region. Let us now assume a very slowly 
varying field according to Eq. (\ref{bprofile}), such that $B_s=8.76\times 10^{15}$ G, $B_{\rm cent}=8.8\times10^{15}$ G, 
$\gamma=0.47$ and $\eta=0.00178$ for 
$E_{Fmax}=20m_ec^2$. If we repeat the same calculation in the Newtonian framework by including the corresponding 
$\rho_B$ and $dP_B/dr$ in the magnetostatic balance and 
mass equations appropriately, we arrive at $M=2.44 M_\odot$ and $R=570$ km, 
practically the same as what we have already reported \cite{prd12,prl13}. Such a super-Chandrasekhar white dwarf would 
have almost zero $dP_B/dr$ compared to $dP/dr$ and hence would be stable. 
Interestingly, neutron stars under magnetar models with $B_s\sim 10^{15}$ G 
are also generally believed to have little difference between the magnetic fields at center and 
surface, when the superconducting effect could not allow a large $B_{\rm cent}$, making 
$dP_B/dr$ very small. This is analogous to the aforementioned example of super-Chandrasekhar 
white dwarfs and, hence, they could be excellent candidates for 
soft gamma-ray repeaters (SGRs) and anomalous X-ray pulsars (AXPs). Nevertheless, the other stable
super-Chandrasekhar white dwarfs, having a smaller $B_s$, can also adequately explain 
SGR/AXP phenomena \cite{mukhrao}.

\acknowledgments

B.M. acknowledges partial support through research Grant No. ISRO/RES/2/367/10-11. 
U.D. thanks CSIR, India for financial support.

\end{document}